\documentclass[aps,prl,twocolumn,superscriptaddress,showpacs]{revtex4-1}
\def\be{\begin{equation}}
\def\ee{\end{equation}}
\def\bea{\begin{eqnarray}}          
\def\eea{\end{eqnarray}}
\def\bi{\begin{itemize}}
\def\ei{\end{itemize}}

\usepackage{tikz}
\usetikzlibrary{arrows,shapes}

\begin{document}

\tikzstyle{every picture}+=[remember picture]

\title{Entanglement Symmetry, Amplitudes, and Probabilities: Inverting Born's Rule}
\date{\today}

\author{Wojciech H. Zurek}
\affiliation{Theory Division, LANL, Los Alamos, NM 87545, USA}

\begin{abstract}
Symmetry of entangled states under a swap of outcomes (``envariance'') implies their equiprobability, and leads to Born's rule $p_k=|\psi_k|^2$. Here I show that the amplitude of a state given by a superposition of sequences of events that share same total count (e.g., $n$ detections of 0 and $m$ of 1 in a spin $\frac 1 2$ measurement) is proportional to the square root of the fraction -- square root of the relative frequency -- of all the equiprobable sequences of 0's and 1's with that $n$ and $m$. 

\end{abstract}

\maketitle

Probability was tied to symmetry since its inception: Laplace \cite{Laplace} used complete ignorance 
-- 
e.g., indifference of the player to shuffling of the deck when face values of the cards are not known -- as evidence of invariance, to define equiprobability. However, the symmetry captured by this ``principle of indifference" is subjective: It does not reflect the state of the deck (shuffling changes order of the cards) but only subjective ignorance of the observer who is unable to predict whether permuting their order will result in a favorable or an unfavorable event. 

Envariant approach to probabilities \cite{Zurek03a,Zurek03b,Zurek05} is based on symmetry -- on the observation that when a perfectly entangled state of any two systems is ``swapped'' on one end, local states of these systems cannot change: Imagine a Bell state $| \heartsuit \rangle_{\cal S} | \diamondsuit \rangle_{\cal E} + |\spadesuit \rangle_{\cal S} | \clubsuit \rangle_{\cal E}$ of ``a system'' and ``an environment'', $\cal S$ and $\cal E$.  
One can use its symmetries to prove that
local state of either is completely unknown. The proof is straightforward: Correlations between possible outcomes in $\cal S$ and $\cal E$ can be manipulated locally. Thus, one can swap $| \heartsuit \rangle_{\cal S}$ and $|\spadesuit \rangle_{\cal S}$ 
by acting only on $\cal S$:

\vspace{.2cm}

\[
\tikz[baseline]{
            \node[fill=gray!20,anchor=base] (t1)
            {$|\heartsuit\rangle_{\cal S}$};
        } 
| \diamondsuit \rangle_{\cal E} 
+
\tikz[baseline]{
            \node[fill=gray!20,anchor=base] (t2)
            {$|\spadesuit \rangle_{\cal S}$};
        }
|\clubsuit \rangle_{\cal E}
\quad \longrightarrow \quad
| \spadesuit\rangle_{\cal S} | \diamondsuit \rangle_{\cal E} + |\heartsuit \rangle_{\cal S} | \clubsuit \rangle_{\cal E}.
\]
\begin{tikzpicture}[overlay]
        \path[->] (t1) edge [bend left=40] (t2);
        \path[->] (t2) edge [bend left=40] (t1);
\end{tikzpicture}

\vspace{-.2cm} 

\noindent Such a unitary {\it swap} exchanges probabilities of the two possible outcomes, $\heartsuit$ and $\spadesuit$ (hence its name). This is obvious, as $\cal E$ 
is untouched by the swap. Therefore, the ``new'' post-swap probabilities of $\heartsuit$ and $\spadesuit$ (that before matched probabilities of $\diamondsuit$ and $\clubsuit$, respectively) must now match the (unchanged) probabilities of $\clubsuit$ and $\diamondsuit$ instead. 

However, global initial state of the whole composite $\cal SE$ 
can be restored by 
a {\it counterswap} in $\cal E$:

\vspace{.2cm}

\[
|\spadesuit\rangle_{\cal S} \tikz[baseline]{
            \node[fill=gray!20,anchor=base] (t3)
            {$| \diamondsuit \rangle_{\cal E}$};
        } 
+
|\heartsuit \rangle_{\cal S}
\tikz[baseline]{
            \node[fill=gray!20,anchor=base] (t4)
            {$|\clubsuit \rangle_{\cal E}$};
        }
\quad \longrightarrow \quad
|\spadesuit \rangle_{\cal S} | \clubsuit \rangle_{\cal E}| + |\heartsuit \rangle_{\cal S} | \diamondsuit \rangle_{\cal E}.
\]
\begin{tikzpicture}[overlay]
        \path[->] (t3) edge [bend left=40] (t4);
        \path[->] (t4) edge [bend left=40] (t3);
\end{tikzpicture}

\vspace{-.2cm}

\noindent This means that probabilities of $\heartsuit$ and $\spadesuit$ are both exchanged (by the swap on $\cal S$) and unchanged (because a (counter)swap in $\cal E$ restores the whole entangled state without touching $\cal S$).
This ``exchanged yet unchanged'' requirement can be met only when the two probabilities are equal, $p_\heartsuit=p_\spadesuit$. With the usual normalization -- certainty 
 corresponding  to the probability of 1 -- we get;
\vspace{-.1cm}
$$p_\heartsuit=p_\spadesuit= 1 / 2 \ .$$ 
Generalization to more dimensions is straightforward.

Note that for general entangled state, e.g. 
$$\alpha | \heartsuit \rangle_{\cal S} | \diamondsuit \rangle_{\cal E} + \beta |\spadesuit \rangle_{\cal S} | \clubsuit \rangle_{\cal E}$$
 with $\alpha\neq \beta$ this proof would fail (as it should). However, envariance is still useful: Swaps are not envariant, but rotation of phases of the coefficients in $\cal S$ by local unitary $e^{-i\varphi}|\heartsuit \rangle_{\cal S} \langle\heartsuit| + |\spadesuit \rangle_{\cal S} \langle \spadesuit |$ is undone by $e^{i\varphi}|\diamondsuit \rangle_{\cal E} \langle\diamondsuit| + |\clubsuit \rangle_{\cal E} \langle\clubsuit  |$.
Thus, envariance implies decoherence -- {\it phases} of the complex coefficients in the {\it Schmidt decomposition} do not matter: Probabilities can depend only on their absolute values \cite{Zurek03a,Zurek03b,Zurek05} -- their amplitudes.

Schmidt decomposition (with {\it orthonormal} partner states in $\cal S$ and $\cal E$) is essential: Local unitaries can alter phases of the coefficients ``one at a time'' only when corresponding states are orthogonal. Moreover, absolute values of the coefficients have any
significance only when states are normalized in the same way. This was key in proving equiprobability. (Overall norm or phase of the entangled state, by contrast, do not matter for us here.)



Envariant proof of equiprobability works for 
equal absolute values of Schmidt coefficients.
The case of unequal coefficients 
 can be always reduced to the case of equal coefficients. To illustrate how, we change notation ($\alpha \rightarrow \sqrt m, \ |\heartsuit\rangle_{\cal S} \rightarrow |0\rangle, \ |\spadesuit \rangle_{\cal S} \rightarrow |1\rangle$, etc.) and consider:
$$|\psi_{\cal SE} \rangle \propto \sqrt n |0\rangle |\varepsilon_0\rangle + \sqrt m |1\rangle |\varepsilon_1\rangle $$
When $n \neq m$ swaps are no longer envariant, as counterswaps do not restore the initial state, $ \sqrt n |0\rangle |\varepsilon_0\rangle + \sqrt m |1\rangle |\varepsilon_1\rangle \neq \sqrt m |0\rangle |\varepsilon_0\rangle + \sqrt n |1\rangle |\varepsilon_1\rangle $. We assume that $m$, $n$ are integers, and that Hilbert space of $\cal E$ is either large enough or can be enlarged to allow for a basis change: 
$|\varepsilon_0\rangle= \frac 1 {\sqrt n} \sum_{i=1}^n |\epsilon_i\rangle, \  |\varepsilon_1\rangle=\frac 1 {\sqrt m} \sum_{i=n+1}^{m+n} |\epsilon_i\rangle $, so that:
\vspace{-0.2cm}
$$|\psi_{\cal SE} \rangle \propto \sqrt n |0\rangle (\frac 1 {\sqrt n} \sum_{i=1}^n |\epsilon_i\rangle) + \sqrt m |1\rangle (\frac 1 {\sqrt m} \sum_{i=n+1}^{m+n} |\epsilon_i\rangle) \ . $$
\vspace{-0.1cm}
In effect, coefficients are replaced by counting;
\vspace{-0.2cm}
$$ |\psi_{\cal SE} \rangle \propto |0\rangle  \sum_{i=1}^n |\epsilon_i\rangle + |1\rangle \sum_{i=n+1}^{m+n} |\epsilon_i\rangle \ . $$
In this state each $|0\rangle|\epsilon_i\rangle$ and each $|1\rangle|\epsilon_i\rangle$ have the same coefficients, so we can again appeal to swaps. It follows that (as a result of deliberate ``finegraining'' $|\varepsilon_0\rangle$ and $|\varepsilon_1\rangle $) the probability of every $|0\rangle|\epsilon_i\rangle$ and $|1\rangle|\epsilon_i\rangle$ is the same:
$$ p_{0,i}=p_{1,i}= \frac 1 {m+n} $$ 
We can prove this using envariance. To this end, we consider an additional $\cal E'$ that entangles with $\cal SE$ so that $|e_i\rangle$, its orthogonal states, become correlated with 
$|\epsilon_i\rangle$:
$$|\Psi_{\cal SEE'} \rangle \propto |0\rangle  \sum_{i=1}^n |\epsilon_i\rangle|e_i\rangle + |1\rangle \sum_{i=n+1}^{m+n} |\epsilon_i\rangle|e_i\rangle \ . \eqno(1)$$
Now a swap of $|e_k\rangle$ with $|e_l\rangle$ can be undone by a counterswap of $|0\rangle   |\epsilon_k\rangle$ with $|1\rangle  |\epsilon_l\rangle$, partner states in $\cal SE$. Thus 
$ p_{0,i}=p_{1,i}= \frac 1 {m+n} $. If we assumed 
additivity of probabilities, Born's rule \cite{Born} with the probabilities proportional to squares of the coefficients in $ |\psi_{\cal SE}\rangle$, would follow:
$$ p(0) = \frac n {m+n} , \ \ \ \ p(1) = \frac m {m+n} \ ; \eqno(2)$$
Envariance under swaps obviates the physical significance of Born's rule. This contrasts with Gleason's measure-theoretic proof \cite{Gleason} which makes no contact with physics. 

Envariance -- deducing equiprobability 
from invariance under swaps -- provides the key physical insight.
Moreover, with envariance one can avoid assuming additivity. (Gleason assumed additivity at the outset.) This is obvious for $m=1$, as then:
$ p(0) = \frac n {n+1}, \  p(1)=\frac 1 {n+1}.$
The first equality follows from the normalization (union of an event and its logical complement is certain, so a sum of their probabilities is 1). Using finite induction that starts with this simple case one can show, {\it without assuming additivity of probabilities}  \cite{Zurek05}, that $ p(0) = \frac n {m+n} , \ p(1) = \frac m {m+n}$, Eq. (2). So, deriving Born's rule does not require additivity. (Additivity is also not needed in the classical equiprobability-based approach \cite{Gnedenko}.) This is important, as in a theory with an overarching additivity principle (quantum principle of superposition) imposing another additivity demand
(that is at odds with the superposition principle, e.g., in double slit experiments) is 
problematic. Envariance of Schmidt coefficients phases -- decoherence pointed out earlier -- is behind this ``emergent additivity''.

The discussion above relied on commensurate (squares of) the coefficients. The incommensurate case is handled \cite{Zurek03a,Zurek03b,Zurek05} by the appropriate limiting argument and the assumption that probability is a continuous function of pre-measurement states. Then, for any state, one always can devise commensurate sequences of states that converge on it and bound probabilities it implies from above and below. This ``Dedekind cut'' strategy is straightforward. Commensurate sequences used to implement it are amenable to envariant treatment described above.

We now have a clear and physically well motivated derivation of Born's rule from the basic ``no-collapse'' principles of quantum theory and from the assumption that probability of a measurement outcomes is continuous in the pre-measurement state.
Our goal is, in a sense, to reverse it. We shall now appeal to symmetries of entangled states to show that 
the {\it amplitude is proportional to the square root of the number of ``favorable outcomes"}. 

Consider a collection of $M$ identical copies of $\cal S$:
$$ |\breve \psi_{\cal S} \rangle = \bigotimes_{k=1}^M (\alpha |0\rangle + \beta|1\rangle)_k  $$
Memory cells of the apparatus $\cal A$ entangle with the system in course of the (pre-)measurement leading to:
$$ |\breve \Psi_{\cal SA} \rangle =  \bigotimes_{k=1}^M (\alpha |0\rangle |a_0\rangle + \beta|1\rangle |a_1\rangle)_k =  \bigotimes_{k=1}^M |\Psi_{\cal SA} \rangle_k \ . \eqno(3a)$$
We could include environment and decoherence, and discuss interactions that correlate outcome states of $\cal S$ with $\cal E$, disseminating measurement result throughout $\cal E$ and making it objective via quantum Darwinism ~\cite{Zurek03b,Schlosshauer,Zurek09}. States in each of $M$ instances 
would have a form:
$$ |\Psi_{\cal SAE} \rangle_k=(\alpha |0\rangle |a_0\rangle \otimes_{i=1}^L |\varepsilon_0 \rangle_i + \beta|1\rangle |a_1\rangle\otimes_{i=1}^L |\varepsilon_1 \rangle_i)_k \ .$$
Such detailed description with multipartite $\cal E$ would only complicate the notation and obscure the essence of what follows. 
We work with the simpler $|\breve \Psi_{\cal SA} \rangle $ representing the whole ensemble. Indeed, one could ``absorb'' the environment by regarding $\cal E$ as a part of $\cal A$, and redefine notation so that $|a_0\rangle =|a_0\rangle \otimes_{i=1}^L |\varepsilon_0 \rangle_i ~, \ |a_1\rangle =|a_1\rangle \otimes_{i=1}^L |\varepsilon_1 \rangle_i $. Whether the reader decides to implement this change of notation will not matter for the remainder of this paper.

As before, we begin with the case of equal coefficients, $\alpha=\beta$. The state vector $|\breve \Psi_{\cal SA} \rangle $ is then envariant under a swap $(|0 \rangle \langle 1 |+|1 \rangle \langle 0 |)_k$ acting on $k$'s member of the ensemble, as such a swap can be undone by a counterswap $(|a_0 \rangle \langle a_1 |+|a_1 \rangle \langle a_0 |)_k$ acting on $\cal A$. This envariance under swaps is preserved when the state of the whole ensemble is  expanded into the sum of the form:
$$|\breve \Psi_{\cal SA} \rangle \propto \sum_{m=0}^M |\tilde s_m\rangle \eqno(3b)$$
where each {\it unnormalized} $|\tilde s_m\rangle$ represents all sequences of outcomes and records that yield $m$ detections of ``1":
\begin{eqnarray}
\tighten
&|&\tilde s_0\rangle  = |00 ... 0\rangle |A_{00 ... 0} \rangle \nonumber  \\ \nonumber
&|&\tilde s_1\rangle  = |10 ... 0\rangle |A_{10 ... 0} \rangle + |01 ... 0\rangle |A_{01 ... 0} \rangle + ...
\\ \nonumber  
& &\ \ \ \ \ \ \ \ \ \ \ \ \ \ \ \ \ \ \ \ \ \ \ \ \ \ \ \ \ \ \ \ \ \ \ \ \ \ \ \ \ \ \ \ \ \ ...+ |00 ... 1\rangle |A_{00 ... 1} \rangle  
\\ \nonumber
&&..... \\ 
\nonumber &|&\tilde s_m\rangle  = |11...1100 ... 0\rangle |A_{11...1100 ... 0} \rangle + ...
 \\ \nonumber 
& &\ \ \ \ \ \ \ \ \ \ \ \ \ \ \ \ \ \ \ \ \ \ \ \ \ \ \ \ \ \ ... + |00...0011 ... 1\rangle |A_{00...0011 ... 1} \rangle \\ \nonumber
&&..... \\
&|&\tilde s_M\rangle  =  |11 ... 1\rangle |A_{11 ... 1} \rangle \nonumber \ \ \ \ \ \ \ \ \ \ \ \ \ \ \ \ \ \ \ \ \ \ \ \ \ \ \ \ \ \ \ \ \ \ \ \ \ \ \ \  (4)
\end{eqnarray}
Above, memory state of the apparatus 
is the product of the 
record states of individual measurement outcomes, e.g. $|A_{10...0} \rangle = |a_1\rangle_1 |a_0\rangle_2...|a_0\rangle_M$. All outcome sequence states and all records sequence states are orthonormal. Thus, writing the state $|\breve \Psi_{\cal SA} \rangle $ as above -- as a sum over sequences of outcome states and corresponding record states -- constitutes its Schmidt decomposition. 

There are $\frac {M!} {m!(M-m)!}$ outcome sequence states in $|\tilde s_m\rangle $. Thus, probability of detecting $m$ 1's is proportional to ${M \choose m}$: This is because every outcome sequence state is equiprobable -- it
can be envariantly swapped with any other outcome sequence state -- and, as noted earlier, phases do not matter.
For instance, $|00 ... 0\rangle$ in $|\tilde s_0 \rangle$ can be swapped with $|10 ... 0\rangle$ in $|\tilde s_1\rangle$. The pre-swap $|\breve \Psi_{\cal SA} \rangle$ 
can be restored by counterswap of the corresponding $|A_{00 ... 0} \rangle$ with $|A_{10 ... 0} \rangle$. When $\alpha=\beta$
relative normalizations of all such sequences are the same. 
It follows that 
every permutation of outcomes has a probability of $2^{-M}$, regardless of the number of 1's. This includes sequences with the unlikely total counts such as $|\tilde s_0\rangle$ and $|\tilde s_M\rangle$. 

Such ``maverick'' sequences were regarded as a threat to 
predictive power of quantum theory in interpretations that rely on purely unitary evolutions \cite{Everett,DeWitt,DeWittG}: Their presence made it impossible to establish Born's rule, as there was no way to relate coefficients of outcome states to their probabilities, so every state in the superposition could even be equally likely. One could get rid of maverick branches by asserting that states with sufficiently small amplitude are impossible for some reason \cite{Hsu}, or let $M=\infty$ (so that ``maverick coefficients'' disappear  \cite{Hartle}) but there are valid concerns  \cite{Stein} about such strategies. Envariance makes it clear why such extraordinary measures are not needed: The numbers of maverick sequences are dwarfed by the equally probable ``run of the mill'' sequences. This argument could not be made earlier, as it uses an independent envariant proof of equiprobability.

We now return to the basic question: How to relate the probability of a specific count of, say, $m$ 1's with the amplitude of the corresponding state. Our discussion has prepared us for this. We address it operationally by adding 
a counter $\cal C$ -- another quantum system (e.g., a special purpose quantum computer) -- that computes the number of 1's in each record sequence of the apparatus:
$$|\breve \Upsilon_{\cal SAC} \rangle \propto \sum_{m=0}^M |\tilde s_m\rangle |c_m\rangle \ . \eqno(5a) $$
Here $|c_m\rangle$ are orthonormal states of $\cal C$ that correspond to the distinct totals. We now apply envariance to $|\breve \Upsilon_{\cal SAC} \rangle $
and use it to deduce probability of a specific count ``$m$''.

To do this we first normalize states $|\tilde s_m\rangle$ in Eq. (5a). 
(Without normalization, amplitudes have no mathematical meaning.) 
This is straightforward: Every individual sequence of 0's and 1's 
has the same norm. Therefore, the number of sequences that yield the total count of $m$ 1's determines norms of 
$|\tilde s_m\rangle$; 
$$\langle \tilde s_m | \tilde s_m \rangle \propto  {M \choose m}
 \ . \eqno(6a) $$
Note that, at this stage, we are just carrying out a {\it mathematical} operation that obtains from $|\tilde s_m\rangle $ the corresponding normalized state that can be legally used to implement the Schmidt decomposition.
As noted earlier, normalization of Schmidt states is essential:
Without normalization absolute values of the coefficients of Schmidt states have no mathematical (or physical) significance. 

It is easy to see that states
$$ | s_m \rangle = 
{M \choose m}^{-\frac 1 2} | \tilde s_m \rangle \eqno(6b)$$
have the same normalization. The state of the whole ensemble amenable to envariant treatment is then:
$$|\breve \Upsilon_{\cal SAC} \rangle \propto \sum_{m=0}^M 
{M \choose m}^{\frac 1 2}  | s_m\rangle |c_m\rangle =  \sum_{m=0}^M \gamma_m | s_m\rangle |c_m\rangle  . \eqno(5b)$$
This is also a Schmidt decomposition, as $| s_m\rangle$ and $ |c_m\rangle$ are orthonormal. Given our previous discussion we already know that the probability $p_m$ of any specific count $m$ is given by the fraction of such sequences. That is:
$$ p_m= 2^{-M} 
{M \choose m} \ . \eqno(6c)
$$
This follows from the 
direct count of the number of envariant (and, hence, equiprobable) permutations of 0's and 1's contributing to $| s_m\rangle$ and, hence, corresponding to $|c_m\rangle$. So, (5b) shows that the amplitude $\gamma_m$ of $|c_m\rangle$ --  of the ``outcome state'' for an observer enquiring about the count of 1's -- is proportional to the number of equiprobable sequences that lead to that count;
$$|\gamma_m| \propto \sqrt { 
M \choose m
}= \sqrt \frac {M!} {m!(M-m)!}\ . \eqno(7)$$
This reasoning ``inverts'' derivation of Born's rule \cite{Zurek03a,Zurek03b,Zurek05}. We have now deduced that absolute values $|\gamma_m|$ of Schmidt coefficients are proportional to the square roots of cardinalities of subsets of $2^{M}$ equiprobable sequences -- states that yield such `$total ~count = m$' composite events.

The crux of the derivation was writing the same global state $|\breve \Upsilon_{\cal SAC}\rangle$ 
as two different Schmidt decompositions,
\begin{eqnarray}
\nonumber |\breve \Upsilon_{\cal S|AC} \rangle & \propto & |00 ... 0\rangle (|A_{00 ... 0} \rangle |c_0\rangle ) 
\\ \nonumber
 & + & |10 ... 0\rangle (|A_{10 ... 0} \rangle|c_1\rangle) + |01 ... 0\rangle (|A_{01 ... 0} \rangle|c_1\rangle )+\nonumber \\ & & \ \ \ \ \ \ \ \ \ \ \ \ \ \ \ \ \ \ \ \ \ \ \ \ \ \ \  ...+|00 ... 1\rangle( |A_{00 ... 1} \rangle |c_1\rangle) \nonumber \\
...... \nonumber \\
\nonumber & + & |11...1100 ... 00\rangle (|A_{11...1100 ... 00} \rangle |c_m\rangle) + 
\nonumber \\ & & \ \ \ \ \ \ \ \ \ \  ...+ |00...0011 ... 11\rangle ( |A_{00...0011 ... 11} \rangle  |c_m\rangle) \nonumber \\ \nonumber
...... \\
& + & |11 ... 1\rangle (|A_{11 ... 1} \rangle |c_M\rangle) \nonumber \ \ \ \ \ \ \ \ \ \ \ \ \ \ \ \ \ \ \ \ \ \ \ \ \ \ \  (8a)
\end{eqnarray}
for the split ${\cal S|AC}$ of the whole into two subsystems, and
\begin{eqnarray}
|\breve \Upsilon_{\cal SA|C} \rangle \propto \sum_{m=0}^M 
{M \choose m}^{-\frac 1 2}  
| s_m\rangle |c_m\rangle =  \sum_{m=0}^M \gamma_m | s_m\rangle |c_m\rangle  \nonumber  \  \ (8b)
\end{eqnarray}
for the alternative ${\cal SA|C}$. Location of the border between the two parts of the whole $\cal SAC$ is the key difference. It redefines ``events of interest''.
The top $|\breve \Upsilon_{\cal S|AC} \rangle$ treats binary sequences of outcomes as ``events of interest", and, by envariance, assigns equal probabilities $2^{-M}$ to each outcome sequence state. By contrast, in $|\breve \Upsilon_{\cal SA|C} \rangle$ the total count $m$ is an ``event of interest'', but now its probability can be deduced from $|\breve \Upsilon_{\cal S|AC} \rangle$, as both 
represent the same state -- the same physical situation. 

The relation of the coefficients of states $|c_m\rangle$ in $|\breve \Upsilon_{\cal SA|C} \rangle$ and equiprobable events in $|\breve \Upsilon_{\cal S|AC} \rangle$ is straightforward: 
States representing composite events are {\it resultant vectors} in the Hilbert space -- superpositions of more elementary events. 
Quadratic dependence of the probability on amplitude reflects ``Euclidean'' nature of Hilbert spaces, where the length of the resultant vector is given by the Pythagorean theorem for orthogonal component states. 

Generalization to 
the case when $\alpha\neq\beta$ 
is conceptually simple. The global state
after the requisite adjustment of relative normalizations 
is then:
\begin{eqnarray}
\small
|\breve \Upsilon_{\cal SAC} \rangle \propto \sum_{m=0}^M  
{M \choose m}^{\frac 1 2}
\alpha^{M-m} \beta^m  | s_m\rangle |c_m\rangle 
= \sum_{m=0}^M  \Gamma_m | s_m\rangle |c_m\rangle  \ . \nonumber
\end{eqnarray}
Coefficients $\Gamma_m$ that multiply $| s_m\rangle |c_m\rangle$ combine on equal footing preexisting amplitudes $\alpha$ and $\beta$ 
from the initial state, 
Eq. (3a), with square roots of Newton's symbols
-- the numbers of corresponding outcome sequences. Once the state representing the whole ensemble is written as $\sum_{m=0}^M  \Gamma_m | s_m\rangle |c_m\rangle$, the origin of the coefficients $\Gamma_m$ (or $\gamma_m$ before) is irrelevant: Observer presented with a state $\sum_{m=0}^M  \Gamma_m | s_m\rangle |c_m\rangle$ and asked to assess probabilities of outcomes $| s_m\rangle|c_m\rangle$ has no reason to delve into combinatorial origins of $\Gamma_m$. For a measurement with outcome states $| s_m\rangle|c_m\rangle$ the origin of the amplitudes $\Gamma_m$ that multiply them is irrelevant. Their absolute values, however, are: Observer could implement envariant derivation ``from scratch", starting with whatever coefficients are there in the initial state, and finegraining (as before, Eq. (1)), to deduce probabilities of various outcomes.  

Derivation of amplitudes of composite events from numbers of equiprobable elementary events turns tables on an old problem. It employs only an ascetic subset of ``textbook'' \cite{Dirac} quantum postulates: (i) {\it States ``live" in Hilbert spaces}; (ii) {\it Evolutions (including measurements) are unitary}. Entanglement is enabled by ``postulate (o)": {\it Hilbert spaces of composite systems have tensor structure}. This is essential for envariance. The need for probabilities is apparent in a ``relative states'' point of view \cite{Everett}, and can be further motivated by the repeatability postulate; (iii) {\it Immediate repetition of a measurement yields the same outcome}. It implies orthogonality of outcomes (or, what is more relevant, of record states $|a_0\rangle,~|A_{0...0}\rangle$, or $|c_m\rangle$) \cite{Zurek07}.  It constitutes a quantum embodiment of ``communicability'' of outcomes emphasized by Bohr \cite{Bohr}. Normalization of outcome states in the Hilbert spaces of $\cal S, \ A$ and $\cal C$ is important. It is a mathematical requirement that 
endows Schmidt coefficients with significance.

Purely quantum ingredients lead to Born's rule \cite{Zurek03a,Zurek03b,Zurek05}. Here we used (o)-(ii) to deduce coefficients of composite event states (total counts $m$) from the numbers of elementary events (detections of ``0'' and ``1''). 
To derive Born's rule from no-collapse quantum postulates we have employed two ideas:  {\it Symmetries of entanglement} establish equiprobability: Envariance was key to our approach.
The second ingredient -- illustrated by Eq. (8a,b) above -- is {\it consistency of amplitude/probability assignments in composite quantum systems}.

Born's rule reflects geometry of Hilbert space. We explored it using swaps and finegraining, but they need not be physically implemented: Postulates (o)-(iii) imply $p_k=|\psi_k|^2$. Nevertheless, it would be extremely interesting to test envariance and finegraining in experiments. It is a very basic and fundamentally quantum symmetry. 

Envariance also relies on locality of quantum dynamics (i.e., the fact that a unitary operation {\it here} cannot change a state {\it there}) and on the basic fact that a state is all quantum theory offers as means of predicting measurement outcomes: Same states imply the same predictions. 

Envariance is an objective property -- a symmetry of entangled states.
Tensor structure of quantum states 
allows for entanglement and for a very different origin of probabilities of a single event 
than subjective ignorance \cite{Laplace}, the sole possibility in classical settings: A perfectly entangled state of the whole can be used to prove rigorously that distinguishabe local states are
envariantly swappable, assuring
equal probabilities to orthogonal outcomes of local measurements. Probabilities in our quantum Universe reflect symmetries of composite systems and are mandated by quantum indeterminacy. Envariance justifies this {\it objective ignorance}.

Envariant derivation of $p_k=|\psi_k|^2$ was by now discussed by others  \cite{Schlosshauer,SchlossFine}. The converse of Born's rule established here is a crucial link, clarifying the relation between quantum states, frequencies, and probabilities.

 
This research was supported in part by DoE through the LDRD grant at Los Alamos.

\end{document}